%% file: main.tex
\documentclass[runningheads]{llncs}
\usepackage[]{graphicx}
\usepackage[]{color}
\usepackage{alltt}
\usepackage[T1]{fontenc}
\usepackage[utf8]{inputenc}
\usepackage{array}
\usepackage{verbatim}
\usepackage{amsmath}
\usepackage{cases}
\usepackage{float}
\usepackage{xspace}
\usepackage{xcolor}
\usepackage[section]{placeins}
\usepackage{adjustbox} 
\usepackage{url}
\usepackage{multirow}
\usepackage{diagbox}
\usepackage{pgfplots} 
\usepackage{lipsum}
\usepackage{mdframed}
\usepackage{caption}
\usepackage{subcaption}
\usepackage{algorithm}
\usepackage{algpseudocode}%
\usepackage{float}

\usepackage{hyperref}
\hypersetup{hidelinks,
            linkcolor = blue} 

\graphicspath{ {images/} }

\begin{document}

\newcommand{\gio}[1]{\textcolor{orange}{\textbf{[GQ] #1}}}
\newcommand{\mer}[1]{\textcolor{cyan}{\textbf{[LB] #1}}}
\newcommand\dyna{\textit{dynaSpark}\xspace}
\newcommand\hyper{\textit{hyperFL}\xspace}
\newcommand\roma{\textit{ROMA}\xspace}

\title{Training and Serving \\Machine Learning Models at Scale}	

\author{Luciano Baresi and Giovanni Quattrocchi}
\authorrunning{L. Baresi and G. Quattrocchi}
\institute{Dipartimento di Elettronica, Informazione e Bioingegneria,
\\ Politecnico di Milano, Italy \\
\email{\{name.surname\}@polimi.it}
}

\maketitle              

\begin{abstract}
In recent years, Web services are becoming more and more intelligent (e.g., in understanding user preferences) thanks to the integration of components that rely on Machine Learning (ML). 
Before users can interact (inference phase) with an ML-based service (ML-Service), the underlying ML model must learn (training phase) from existing data, a process that requires long-lasting batch computations. The management of these two, diverse phases is complex and meeting time and quality requirements can hardly be done with manual approaches.

This paper highlights some of the major issues in managing ML-services in both training and inference modes and presents some initial solutions that are able to meet set requirements with minimum user inputs. A preliminary evaluation demonstrates that our solutions allow these systems to become more efficient and predictable with respect to their response time and accuracy.
\end{abstract}


\section{Introduction}\label{sec:intro}
\input{introduction}

\section{ML-service Management}\label{sec:problem}
\input{problem}

\section{What we have done so far}
\label{sec:solution}
\input{solution}

\section{Conclusions}
\label{sec:conclusions}
\input{conclusions}

\bibliographystyle{plain}
\bibliography{referencelist}

\end{document}

%% file: introduction.tex
In recent years, Web services have been enriched with components based on  Machine Learning (ML) that allow for new types of interactions (e.g., vocal assistants, chatbots), more customized experiences (e.g., recommender systems), and novel features (e.g., object detection in images, smart auto-completion)~\cite{jordan2015machine}. ML-based services (ML-Services) exploit models that,  created in a preliminary \textit{training} phase, are then used at runtime to provide required results as predictions on new inputs (\textit{inference} phase) \cite{mohri2018foundations}. 

The training phase is a long-lasting ---from several minutes to days--- batch computation. Given the inputs of this phase are usually large datasets (e.g., thousands of images), the computation is executed by using dedicated frameworks (e.g., Spark~\cite{spark}, TensorFlow~\cite{tensorflow}) that run on a highly distributed cluster of virtual or physical machines. 
The inference phase exploits the generated model in an interactive way and each computation (e.g., a prediction based on a single image as input) lasts a significantly shorter amount of time compared to 
training (e.g., milliseconds or seconds). 

The quality of these ``new'' web services is not only stated in terms of constraints on provided performance. For example, traditional Service Level Agreements (SLA) or Service Level Objectives (SLO)~\cite{sahai2002towards} usually constrain the maximum allowed response time. An ML-service must also be accurate. It is not a matter of functional correctness (and performance) anymore; an ML-service must offer predictions with a given accuracy and precision~\cite{juba2019precision}.

In the training phase, requirements on the response time are specified as \textit{deadlines}~\cite{islam2020cost}, which is the maximum allowed time to complete a single training process. In the inference phase, the response time of multiple requests (e.g., the ones received in the last 10 minutes) is aggregated and constrained by a threshold~\cite{labidi2017cloud}. 
In the training phase, requirements on quality can be easily defined through thresholds (e.g., the accuracy should be greater than $90\%$) since this metric can be measured during the computation using a validation or a testing dataset~\cite{vabalas2019machine}. However, in inference mode, the quality of the predictions cannot be easily computed since, by definition, the correct output is unknown (if it were not, there would not have been any need for the ML model). Thus, different ``indirect'' metrics can be considered to estimate the quality of predictions at runtime~\cite{weiss2021uncertainty}.

The management of ML-services is thus a complex task since multiple interdependent factors affect their overall performance. For example, the response time is affected, among other aspects, by resource allocation. ML computations can be executed on CPUs, but also on dedicated hardware (e.g., GPUs) given their highly parallelizable programming model. However, increasing the computational power is not useful if the data or the model itself are not properly partitioned to allow for a sufficient degree of parallelism. Model quality depends on a number of factors including the parameters of the learning algorithm (i.e., hyperparameters) and the characteristics of the input dataset. Moreover, response time and quality are often correlated. For example, fewer learning iterations allow for faster results but produce, in general, less accurate models.

Some initial approaches have already been presented to tackle single aspects of ML-service management. For example, Nguyen et al.~\cite{Nguyen2018TowardsAT} present a solution to predict the time required to train a model using Spark given a user-defined configuration. Li et al.~\cite{li2020system} describe a solution for the automated configuration of hyperparameters, while Morabito et al.~\cite{morabito2021discover} propose a solution for the automated resource provisioning of ML models in inference mode. However, there is no solution or study that provides a multi-faceted view of the problem.

This paper describes the main aspects and issues of managing ML-services in both training and inference modes. Based on our work in the field, we also illustrate a set of solutions, along with their evaluation, that we have developed during the last few years and that can be used as a starting point to design a fully integrated solution for the management of ML-based systems. 

The rest of this paper is organized as follows. Section~\ref{sec:problem} presents the most important challenges of ML-services. Section~\ref{sec:solution} describes our solutions and reports their initial evaluation. Finally, Section~\ref{sec:conclusions} concludes the paper.

%% file: problem.tex
This section highlights the main aspects and issues related to ML-service management. It mainly focuses on time and quality constraints during both training and inference phases. 

\subsection{Training}

Training computations require executing highly iterative algorithms that exploit large datasets as inputs. These datasets contain known input-output pairs that are used to generate models that can predict the correct outputs for new, never-seen inputs (inference). 

The structure of the model (e.g., the layers of a neural network) is defined by the users beforehand along with a set of \textit{model parameters} whose optimal values are computed as a result of the training process.
To do so, the training phase minimizes a \textit{loss} function that compares the outputs generated by the model under construction and the known answers. 

During the optimization process, the model parameters are updated multiple times. The execution is organized in \textit{batches}. A batch is the amount of data (i.e., the number of input-output pairs) that is used to update the model parameters. An \textit{epoch} is the amount of batches required to process the whole dataset. Each training process executes multiple epochs, up until a certain model quality or a timeout are reached.

\subsubsection{Available frameworks}
Given that these computations are resource-demanding and require processing large amounts of data that cannot easily fit on a single machine, the training phase of ML models is usually executed through special purpose frameworks that allow for highly distributed and parallelized executions.

The reference programming model for batch computations is \textit{map-reduce}~\cite{mapreduce}, popularized by Google and the Hadoop framework~\cite{hadoop}. The computation is organized in two phases: map and reduce. In the former, the input dataset (e.g., a set of user posts) is split into a set of partitions that are processed in parallel by multiple executors that transform each partition into a new set of data (e.g., a list of words occurrences). The latter aggregates the transformed data and produces a result (e.g., the top ten words used by users in their posts) that is usually written into a database or disk. 

\textit{DAG-based} frameworks, such as Spark, are an evolution of this programming model and they allow one to create \textit{direct acyclic graphs} (DAGs) of data transformations and aggregations (not only a sequence of map-reduce computations). As in map-reduce, multiple executors process single partitions of data in parallel. These computations are organized in stages, that is, the execution of a set of operations that do not require data transmission among executors. When a stage is concluded, each executor sends the results of its local computation to another executor (data shuffling), and a new stage (the next in the DAG) is scheduled for execution. In these frameworks, computations and data transfers are executed in-memory without requiring to store intermediate results on (slower) external storage systems.

The aforementioned approaches are general-purpose and not specifically designed for ML training.
In recent years, new frameworks, such as TensorFlow~\cite{tensorflow} and PyTorch~\cite{pytorch}, have been developed as highly optimized frameworks for ML training. They introduce a programming model called \textit{all-reduce} where the executors work on a single batch in parallel with the others. The computed results are then aggregated (\textit{reduce}) and the model update is transmitted to all the other executors (\textit{all}) so that they can use it as a starting point for the next batch. This way, all-reduce allows for the efficient propagation of the results of an aggregation to all the available executors (broadcast), whereas in DAG-based computations a single executor only communicates with another one and broadcast communication is less efficient and more difficult to implement.

All these frameworks do not provide explicit means to set time or quality requirements. Therefore, users must rely on manual solutions and they have to use experience and ``guess'' the proper configuration that meets expected thresholds.

\begin{mdframed}
 Available frameworks provide different programming models. All of them organize the computation in phases (e.g., one phase for each epoch or each batch) to orchestrate multiple distributed executors that work in parallel. None of these frameworks allows one to specify quality and time constrains on carried out activities.
\end{mdframed}

\subsubsection{Hyperparameters} 

While parameters usually refer to the variables optimized during the training process (model parameters), hyperparameters are the variables used to configure the training algorithm beforehand. These parameters deeply affect both the quality of the output and the time required to complete the training process.

The architecture of the model employed for training can be seen as a hyperparameter. In general terms, the more sophisticated the architecture is, the more time is needed to train the model. In terms of accuracy, complex architectures may be able to learn more patterns and subtleties from the data but they may fail in generalizing to new inputs (by assuming that the subtle patterns found in the input dataset characterize any inputs). 

Any configuration parameter of the training algorithm is a hyperparameter. For example, the batch size regulates the frequency of updates. Frequent model updates may generate more accurate models but also increase the required synchronization among the executors (e.g., higher response times). The number of epochs regulates the amount of iterations on the entire dataset. The more epochs one can afford, the more optimized model parameters will be. This produces high-quality models at the cost of longer executions.

\begin{mdframed}
 Hyperparameters are defined at design time and they impact the quality of the  model and the execution time of the training phase. They must be carefully tuned to train the model efficiently and precisely.
\end{mdframed}

\subsubsection{Parallelism}
To allow multiple executors to cooperate in training models, one can adopt two main strategies to maximize the degree of parallelism of the computations~\cite{jia2019beyond}: \textit{data parallelism} and \textit{model parallelism}.

\textit{Data parallelism} allows one to split the input dataset onto multiple executors, which store a full copy of the model each. This way, the executors can work in parallel on different partitions and the results of the local computations are aggregated in a subsequent phase. This technique is valuable when the input dataset exceeds the memory (and/or storage) of a single executor and allows us to speed up the computation by increasing the number of executors (up to the number of batches).
This strategy is available in most of the existing frameworks and it is the easiest to implement.

\textit{Model parallelism} partitions the model architecture into different parts that are independently stored onto different executors. This strategy is used when the model itself exceeds the capacity of the executor. Moreover, the model architecture may allow for an ``intrinsic'' parallelism so that some of the model parameters can be optimized in parallel with others. In this case, it is very important to understand the best way to partition the model given the executors available. Recent studies show that model parallelism may also slightly decrease model accuracy~\cite{chen2018efficient}.

Data and model parallelism can be combined so that both the model and data are partitioned into multiple parts. Each model partition is then replicated onto multiple executors so that they can optimize a subset of model parameters on independent partitions of data. 

\begin{mdframed}
 Data and model parallelisms allow for highly distributed computations. By partitioning datasets and/or models one can accelerate the computation at the cost of more synchronization among executors. 
\end{mdframed}

\subsubsection{Resource Allocation}
To speed up (or slow down) the training phase and meet set deadlines, users can tune the amount of resources allocated to the system. The computations to train ML models are mostly based on matrix operations that can be highly parallelized. For this reason, GPUs and dedicated hardware devices (e.g., TPUs~\cite{jouppi2018domain}) can be exploited to further accelerate the processing. 
Resource allocation can be handled at both design time (before training) and runtime (during training).

At design time the number of executors and the resources allocated to them must be configured by considering set requirements, the hyperparameters (e.g., number of epochs), and the characteristics of the input dataset. 
More sophisticated approaches allow for dynamic resource allocation (e.g., as in Spark\footnote{\url{https://spark.apache.org/docs/latest/job-scheduling.html}}. This means that the amount of executors and their resource allocation can be re-configured over time according to the state of the computation. For example, when the set deadline is approaching, the system may spawn a new executor or increase the amount of computational power allocated to an existing one. Similarly, one can realize that the computation could require more iterations than expected to reach a set model quality and more resources could help finish the training process in time. 

Memory is also extremely important since slow read/write on the disk can quickly become the bottleneck. The memory allocation of each executor should be adequate to fit both data and model (partitions). Unlike computational power, which can be increased to speed up the processing up to the maximum degree of parallelism, memory is either sufficient or insufficient. 

\begin{mdframed}
Resource allocation is key to fulfill time-based requirements. Resource allocation can be tuned at either design time or runtime and executors may be equipped with heterogeneous hardware. 
\end{mdframed}

\subsection{Inference}

Once training activities have generated a model, it is deployed and used to compute predictions on new inputs. These computations are interactive and they are much faster than training: users can submit requests and get a response in the order of seconds or milliseconds. 

Unlike training where requirements are defined for the complete process (deadline), during inference, they constrain a set of requests received by the system during a given time window. Requirements can be both related to time (e.g., the 95th percentile of the response time should be less than one second) or to quality.

\subsubsection{Available frameworks}

While map-reduce and DAG-based frameworks were designed to host batch computations only, ML frameworks provide some tools to serve models in inference mode. For example, TensorFlow provides \textit{TensorFlow serving}\footnote{\url{https://www.tensorflow.org/tfx/guide/serving}}, a tool that eases the deployment of models onto web servers to let them be invoked by using a REST API. 

When the incoming workload cannot be handled by a single executor, multiple replicas of the same model must be used concurrently (\textit{distributed inference}). Ideally, a framework should allow one to submit requests to a unified API that forwards them to available replicas. Frameworks should also support multiple models running in parallel on a shared infrastructure, since a single ML-service may exploit multiple ML-based components.
Unfortunately, in the available frameworks, the support for distributed inference is, at the time of writing, very limited.  Moreover, none of the frameworks support time or quality requirements, and users must handle them with external or manual solutions. 

\begin{mdframed}
 ML frameworks allow one to use models in inference mode but the means to deploy and manage them are limited. Existing frameworks lack support for the specification of quality and time requirements.
\end{mdframed}

\subsubsection{Resource allocation}
In inference, resource allocation is key to be able to keep the ML-service always responsive. Static resource allocation makes only sense when the workload is constant, a scenario that is usually not realistic.  When the incoming workload fluctuates, resources must be dynamically provisioned to handle the traffic. Executors can be replicated or reconfigured through CPUs, GPUs, or dedicated hardware allocations. 

When a new request is submitted, it must be processed by an available executor. The selection of this executor may be decided according to multiple factors, including the equipped hardware and the amount of requests that it is already handling (queue length). If an ML-based component is close to violating a requirement (e.g., the response time is increasing), faster executors must be selected (e.g., ones equipped with GPUs). If the ML-service is stable, one can leave faster executors to other components.

Memory must be large enough to contain the model, whereas input data are usually much smaller (e.g., a single image) than the ones used in training and they are usually not difficult to handle. If the model is too large to fit onto a single executor, model partitioning can be used to further split the computation.

\begin{mdframed}
Dynamic resource allocation allows for keeping ML-service always responsive when the incoming workload fluctuates. At runtime, requests must be scheduled to proper executors according to their hardware capabilities and the state of the system.
\end{mdframed}

\subsubsection{Monitoring}
Monitoring the quality of the model in inference mode is a complex task since evaluating predictions on new inputs would require knowing the correct outputs (ground truth). 

Since quality cannot be directly computed, alternative metrics must be taken into account.  \textit{Uncertainty} is a widespread metric used in the literature to estimate the quality of a trained models~\cite{kabir2018neural}. Intuitively, when the uncertainty in predictions is high, it is more likely to produce incorrect outputs. 
For example, one can use a set of similar ML models (e.g., DeepEnsemble~\cite{lakshminarayanan2017simple}) to compute multiple predictions for each input. If the outputs are similar (low variance), the outputs are probably correct. If the outputs are different (high variance), the uncertainty is high and the prediction quality may have dropped. 

In this context, users may set a requirement on the maximum allowed uncertainty and when this threshold is violated, the ML model must be re-trained. This may require the manual collection of a new training dataset or the generation of a new one automatically (e.g., using data augmentation~\cite{shorten2019survey}).

\begin{mdframed}
The quality of ML models at runtime cannot be directly monitored for the lack of ground truth. Uncertainty can help estimate the quality of the model and understand when it must be re-trained.
\end{mdframed}

%% file: solution.tex
This section describes three solutions we developed for managing some aspects of ML-services: \dyna, \hyper, and \roma.  \dyna extends Spark by allowing one to associate time requirements with batch computations, and by means of a sophisticated dynamic resource allocation mechanism. \hyper extends Tensorflow and it is also dedicated to the training phase. It exploits heuristics to set hyperparameters values so that quality constraints are met. Finally, \roma is dedicated to the inference phase. It also extends  TensorFlow and allows one to set response time requirements that are met through the allocation of both CPUs and GPUs.

\subsection{\dyna}

\dyna~\cite{baresi2019fine,baresi2019symbolic} extends Spark by introducing advanced and automated resource management.
\dyna allows users to define deadlines that are considered as the desired response time for a single batch execution (e.g., a training process). The goal of \dyna is to control the resources allocated to the computation so that its execution time is as close as possible to the user-defined deadline. The rationale behind it is that the closer the response time is to the deadline the more efficient the usage of resources is: finishing before the deadline would mean allocating more resources than needed, while terminating afterward implies violating the deadline because of too scarce resource allocation. Moreover, \dyna can manage multiple Spark computations at the same time (e.g., the training phase of different ML models that run concurrently) and keep their execution time under control.

\dyna requires a profiling phase to retrieve the DAG of the computation and performance data of each stage.  During the execution, \dyna exploits dynamic vertical scalability of resources. This means that the resources allocated to executors is continuously reconfigured without the need for restarting them or creating new ones. The framework wraps executors in lightweight containers (e.g., Docker\footnote{\url{https://docker.com}}) and control-theoretical planners are used to compute resource allocation in a fast and fine-grained fashion. 

\dyna  exploits a hierarchical control loop. At the top of the hierarchy, a \textit{memory controller} is in charge of dynamically resizing the amount of memory allocated to each running executor. This controller distributes available memory fairly to all running computations and it is only activated when the user submits a new computation or when one finishes.
When a new stage of a computation starts, a \textit{stage controller} exploits the deadline submitted by the user and the profiling data retrieved beforehand. As a result, this controller computes a local deadline for the stage along with the number of executors needed to fulfill it.
Each executor in \dyna is equipped with an \textit{executor level controller} based on control theory in charge of keeping the execution time as close as possible to the deadline computed for the stage. By exploiting a feedback loop, it monitors the progress of the stage (i.e., how many data samples are processed over the total assigned ones) and allocates processing power (i.e., fractions of CPUs) accordingly. Multiple executors work in parallel on a single stage to fulfill the same local deadline, and data are fairly partitioned among them. This way, the controllers can operate without synchronization and independently from one another.

Since multiple executors can be run on the same node, \dyna exploits a \textit{node level controller} to manage resource contention. This controller collects all the resource allocations computed by the executor level controllers that are running on the same node. If their sum exceeds the capacity of the node, allocations are scaled down according to different strategies such as Earliest Deadline First~\cite{edf} (EDF) or proportionally.

\subsubsection{Evaluation}

\begin{figure}[htbp]
	\centering
	\includegraphics[width=0.82\columnwidth]{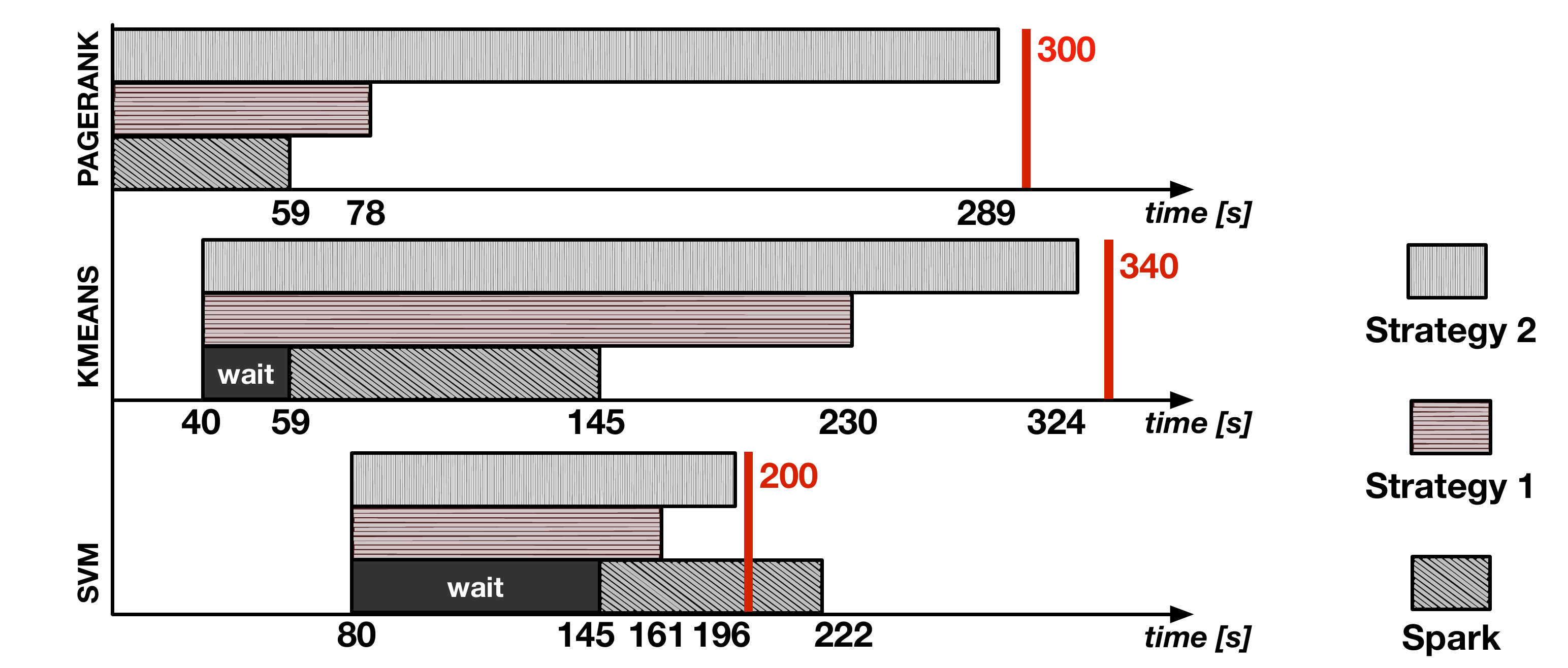}
	\caption{Concurrent computations in \dyna.}
	\label{fig:vsSpark_bench}
\end{figure}

The evaluation of \dyna was executed on Microsoft Azure using a cluster of $5$  \textit{Standard\_D14\_v2} virtual machines (VMs) equipped with 16 CPUs, 112 GB of memory, and 800 GB of local SSD storage. 
Here we only report a single significant experiment that clearly demonstrates the benefits of \dyna, while \cite{baresi2019fine}  comprises a larger and comprehensive set of experiments. 

For this experiment, we compared Spark and \dyna. Spark was configured to use its default allocation mechanism that allocates all the resources to the first computation it receives (FIFO). \dyna was configured with both a resource contention strategy based on EDF (\textit{Strategy 1}) and a proportional one (\textit{Strategy 2}). The first strategy aims to avoid violations and it is more conservative. The second one tends to let the different application progress simultaneously at a similar pace to minimize resources but with more risk of violating the deadlines.

We executed three well-known ML algorithms in parallel: \textit{PageRank}, \textit{KMeans} and \textit{SVM}. As shown in Figure~\ref{fig:vsSpark_bench} we submitted for  execution \textit{PageRank} first at instant $0$ with a deadline of $300$ seconds, then \textit{KMeans} at instant $40$ with a deadline of $300$ seconds and, finally, \textit{SVM} at instant $80$ with deadline $120$ seconds.
Results show that Spark violated the deadline of SVM while the executions of PageRank and KMeans finished way before expected. This inefficient behavior is caused by the limited resource allocation mechanism built in Spark, its simple FIFO-based strategy for managing resource contention, and the lack of support for deadlines.

In contrast, \dyna did not violate the deadlines both with \textit{Strategy 1} and \textit{2}. The former always finished significantly ahead of the time given its conservative policy (more resources are allocated).  \textit{Strategy 2} resulted, in this case, in a highly efficient resource usage since the three computations were completed just a few seconds before their respective deadlines. 

\subsection{\hyper}

\hyper~\cite{hyper} is an extension of TensorFlow to allow a set of federated nodes (i.e., Federated ML~\cite{yang2019flconceptapplications}) to cooperatively train an ML model under quality constraints. Since executors are geographically distributed (e.g., running on a set of edge nodes), the cost of synchronization is particularly significant and the amount of communication between executors should be minimal. For this reason, \hyper assumes executors to work on different datasets and models that are trained locally and aggregated only when needed. In particular, the computation is organized in $R$ rounds ($R$ is fixed and given). A round is a sequence of consecutive epochs where the executors perform independently and in parallel the training processes and only merge the results at the end (similarly to stages in Spark).

At the core of \hyper lays an algorithm, executed at the beginning of each round $r$, that takes as input a constraint on model accuracy ($AC$) (e.g., $AC > AC_{SLA}$) and computes the hyperparameter $E^r$ that is the number of epochs to be processed on each executor during $r$ to obtained the desired quality at the end of $R$ rounds. 

The algorithm exploits two alternative heuristics: linear and quadratic interpolation, and works in a black-box way since it does not require any prior information om the model: it only exploits monitored data (e.g., the accuracy of the model after each round). To be properly initialized the algorithm needs two rounds in which the computation is executed using a fixed configuration.
As a first step, \hyper computes the accuracy to be reached at the end of round $r$ ($AC^r$). To do so it assumes either a linear progress towards $AC_{SLA}$ from round $0$ to $R$, or a smoother quadratic one. In the second step, \hyper computes $E^r$ as a function of $AC^r$,  the cumulative sums of the epochs computed in rounds $r-1$ and $r-2$, and the accuracy obtained at the same previous rounds.

\subsubsection{Evaluation}

We evaluated \hyper on a bare-metal single-user server equipped with an AMD Ryzen 5 2600 @ 3.40GHz (6 Cores / 12 Threads) CPU and 32GB DDR4 @3200MHz of RAM running Ubuntu 19.10. We used two real-world ML-services: \textit{MNIST}~\cite{lecun2010mnist} and \textit{Fashion-MNIST}~\cite{DBLP:journals/corr/abs-1708-07747}. The former takes as input an image with a handwritten digit and outputs the corresponding number. The latter receives a gray-scale image of a fashion item and outputs the corresponding class (e.g., t-shirt, bag). 

\begin{table}[t]
\begin{minipage}[b]{0.49\linewidth}
\centering
\begin{tabular}{c|lclc|lclc|}
 & \multicolumn{4}{c|}{MNIST} & \multicolumn{4}{c|}{Fashion-MNIST} \\
 & \multicolumn{2}{c}{Linear} & \multicolumn{2}{c|}{Quadratic} & \multicolumn{2}{c}{Linear} & \multicolumn{2}{c|}{Quadratic} \\
$r$ & $E^r$ & $AC^r$ & $E^r$ & $AC^r$ & $E^r$ & $AC^r$ & $E^r$ & $AC^r$ \\ \hline
1 & 1 & 0.25 & 1 & 0.16 & 1 & 0.22 & 1 & 0.14 \\
2 & 1 & 0.30 & 1 & 0.19 & 1 & 0.30 & 1 & 0.32 \\
3 & 1 & 0.36 & 7 & 0.51 & 1 & 0.38 & 1 & 0.42 \\
4 & 1 & 0.39 & 1 & 0.54 & 1 & 0.45 & 1 & 0.46 \\
5 & 2 & 0.47 & 3 & 0.62 & 1 & 0.51 & 2 & 0.53 \\
6 & 2 & 0.52 & 3 & 0.68 & 1 & 0.57 & 2 & 0.58 \\
7 & 4 & 0.58 & 3 & 0.73 & 1 & 0.60 & 3 & 0.62 \\
8 & 6 & 0.67 & 4 & 0.75 & 1 & 0.62 & 4 & 0.65 \\
9 & 8 & 0.75 & 11 & 0.79 & 4 & 0.67 & 6 & 0.67 \\
10 & 4 & 0.77 & 1 & 0.80 & 2 & 0.68 & 11 & 0.71
\end{tabular}
\caption{\hyper results.}
\label{tab:eval_mnist_fashion_mnist_tr}
\end{minipage}\hfill
\begin{minipage}[b]{0.49\linewidth}
\centering
\includegraphics[width=\columnwidth]{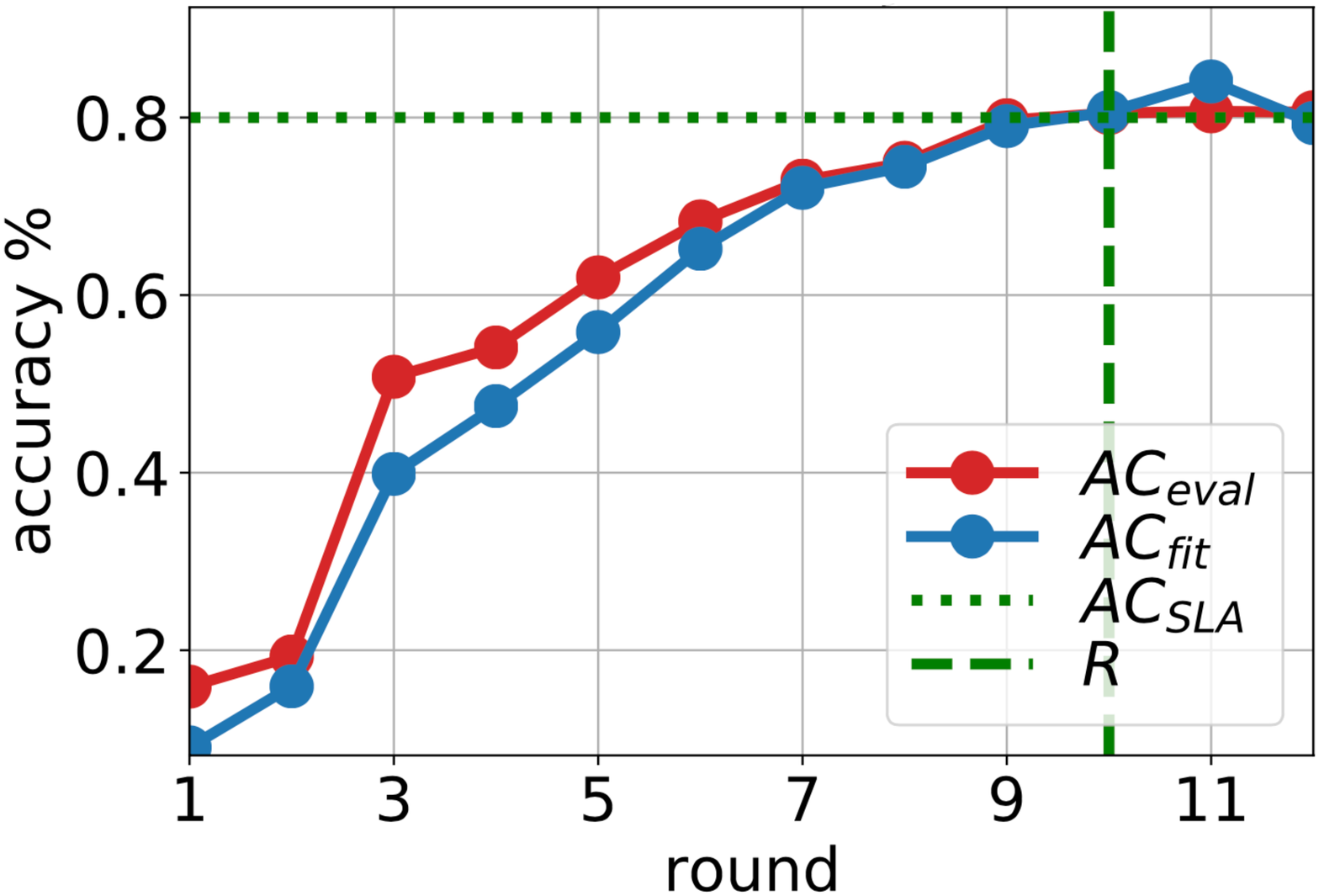}
\captionof{figure}{\hyper accuracy with MNIST.}
\label{fig:accuracy_hyper}
\end{minipage}
\end{table}

In the sample of the experiments reported here, we set for \textit{MNIST} $R=10$ and $AC_{SLA} = 0.80$, while $R=10$ and $AC_{SLA} = 0.70$ for  \textit{Fashion-MNIST}.
Table \ref{tab:eval_mnist_fashion_mnist_tr} shows the computed epoch $E^r$ along with the obtained accuracy $AC^r$ for each round $r$ for both the services and \hyper interpolation algorithms. \hyper with linear interpolation obtained a steady increment of the accuracy with only one epoch computed for rounds 1-4 (\textit{MNIST}) and 1-8 (\textit{Fashion-MNIST}). As final results in both the services, \hyper with linear interpolation terminated the training with an accuracy that is very close to the threshold but slightly lower.

In contrast, \hyper with quadratic interpolation was able in both cases to reach the target $AC_{SLA}$ by computing higher values of $E^r$ in the first rounds. This can be better visualized in Figure~\ref{fig:accuracy_hyper} that reports the result of the experiments with \textit{MNIST} and quadratic interpolation. The chart shows both the accuracy computed on the validation set ($AC_{fit}$) and the one on the testing set ($AC_{eval}$). $AC_{SLA}$ was reached exactly at round $R$, with a peak increment of epochs ($11$) at round $9$. This larger value of $E^r$ corresponds to a small increment in the accuracy ($0.01$) and shows that a quadratic curve better captures the intrinsically asymptotic behavior of the accuracy compared to a linear approximation.

\subsection{\roma}

\roma~\cite{baresi2021resource} is a comprehensive resource management solution for ML models in inference mode. In particular, it facilitates the deployment and operations of multiple interactive ML services on shared infrastructures. 
It extends TensorFlow (and TensorFlow serving) by allowing users to set requirements on the response time. Unlike TensorFlow, \roma provides means for distributed inference and efficiently allocates CPU and GPU resources by considering the state of all served ML-services. 

\roma features a master-slave architecture. The master allows users to submit requests to deployed models and schedules them on the executors that are running on the slaves. Each slave is assumed to be equipped with at least one CPU and zero or more GPUs.  The framework wraps each executor in a container, and multiple replicas of the same model can be deployed when the workload increases. We also exploit Kubernetes\footnote{\url{https://kubernetes.io}} to orchestrate and configure containers. 
The master comprises a \textit{gateway} that store requests ---received through a REST API--- in a dedicated queue (one per ML-service). Two schedulers based on heuristics remove requests from the queues and assign them for execution on a selected executor. A scheduler is dedicated to GPU executions and, as soon a GPU is idle, it extracts a request from the queue of the ML-service that is more likely to violate the set response time. The other scheduler submits requests in a round-robin fashion to available CPUs. 

Each slave can host multiple executors that are controlled by control-theore\-tical planners. As in \dyna, these controllers vertically scale the CPU resources of each container/executor according to the needs of the ML-service. To avoid unpredictable behaviors, control-theoretical planners are aware of GPU executions that can rapidly decrease the response time of the ML-service under control. A supervisor on each slave is in charge of collecting all computed allocations and resizing them if they exceed the capacity of the node.

\roma expects executors to be created manually or automatically using external solutions (e.g. Horizontal Pod Autoscaler\footnote{\url{https://kubernetes.io/docs/tasks/run-application/horizontal-pod-autoscale/}}) and only manages existing resources.
Each executor/container is bound to a specific hardware device. In particular, given $m$ ML-services to be deployed onto a slave node, \roma provisions i) $m$ executors containing one model each, and binds them to the CPUs of the node, and ii) one executor, containing all the models, for each GPU.

\subsubsection{Evaluation}

To evaluate \roma, we used a cluster of three VMs on  Microsoft Azure: one VM of type $HB60rs$  with a CPU with $60$ cores and $240$GB of memory for the \textit{master}, and two VMs, as \textit{slave} nodes, of type $NV6$ equipped with an NVIDIA Tesla $M60$ GPU and a CPU with $6$ cores and $56$GB of memory. We also used an additional $HB60rs$ VM to generate the workloads. 
We exploited four existing ML-services: \textit{Skyline Extraction}~\cite{skyline_extraction},  \textit{ResNet}~\cite{he2016deep}, \textit{GoogLeNet}~\cite{szegedy2015going}, and \textit{VGG16}~\cite{zhang2015accelerating}.  The first service uses computer-vision algorithms to extract the skyline horizon from an input image. The other services perform classification task: \textit{ResNet} uses a residual neural network, while \textit{GoogLeNet} and \textit{VGG16} exploit two  deep convolutional neural networks. 

In the experiments, we run different combinations of these services in parallel under different workloads and we compared the performance of \roma against competitors based on rules or heuristics. The results show that, overall, \roma reduces by $75\%$ on average the number of violations while decreasing by $24\%$ the resources used. 

\begin{figure}[thbp]
	\centering
	\begin{subfigure}[b]{0.49\columnwidth}
		\includegraphics[width=\textwidth]{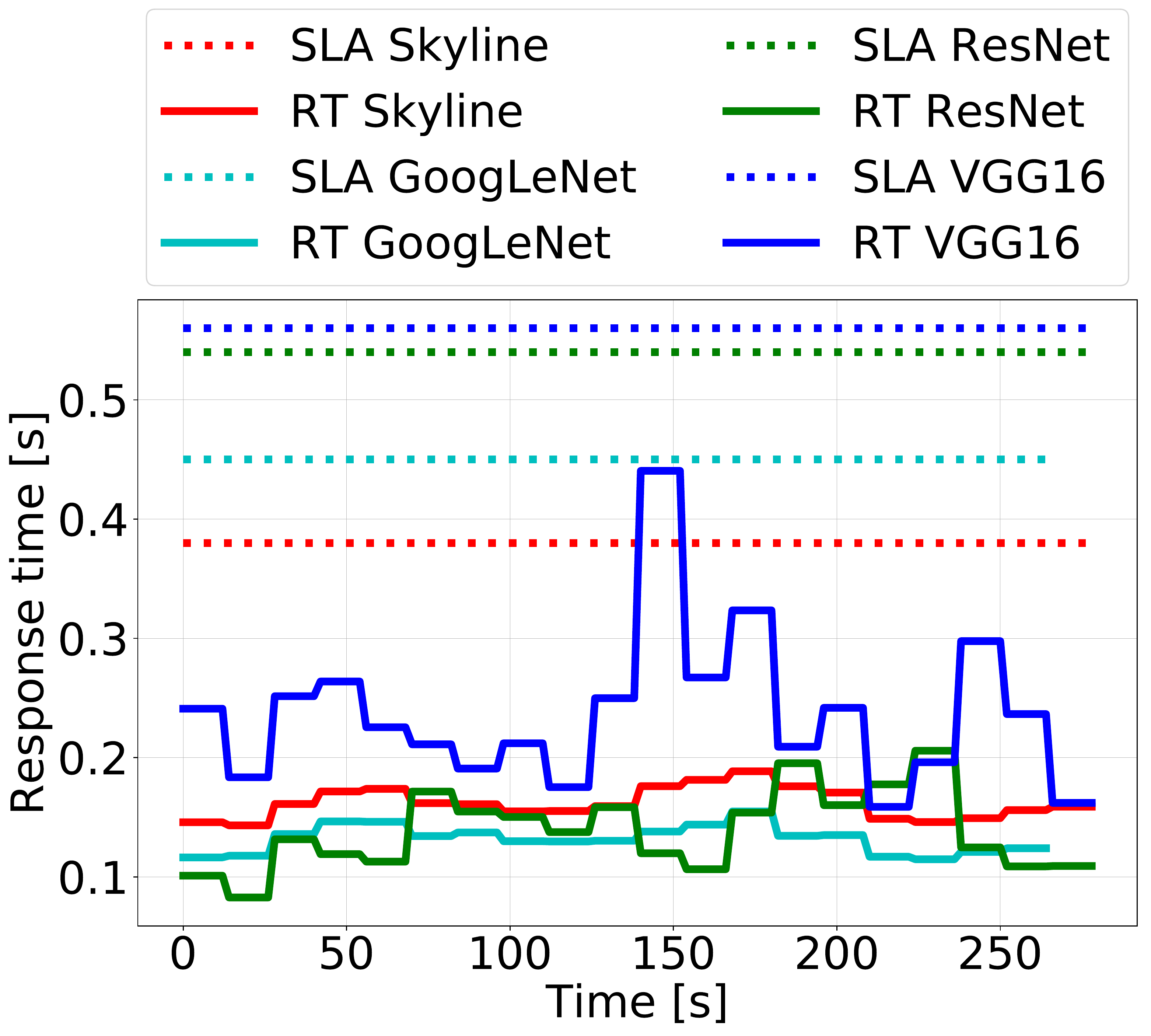}
		\caption{\roma}
		\label{fig:4apps-rt-approach}
	\end{subfigure}
	\hfill
	\begin{subfigure}[b]{0.49\columnwidth}
		\includegraphics[width=\textwidth]{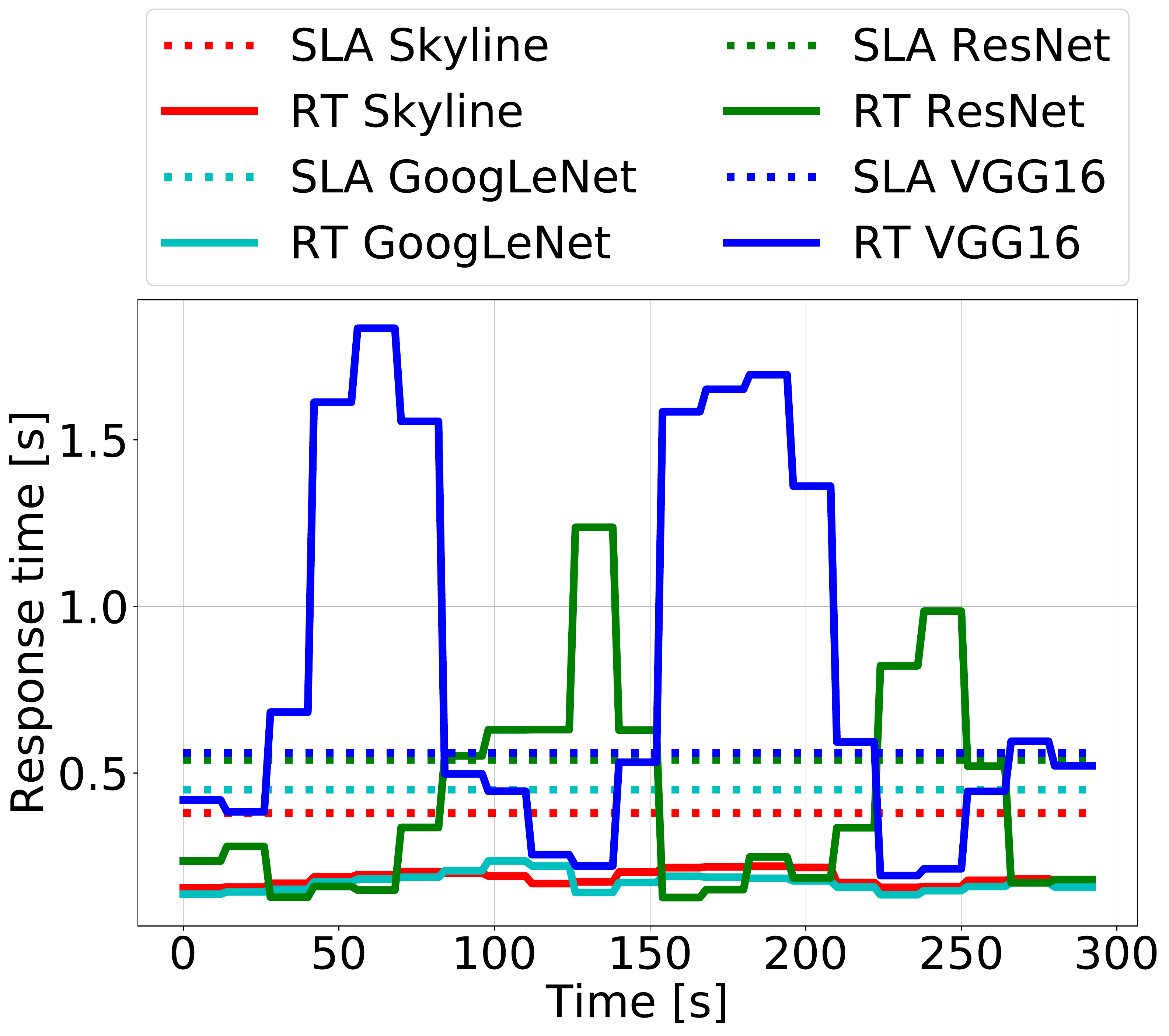}
		\caption{Rule-based approach}
		\label{fig:4apps-rt-rules}
	\end{subfigure}
	\caption{\roma: performance and comparison.}
	\label{fig:roma-comparison}
\end{figure}

Figure~\ref{fig:roma-comparison} shows the most complex experiment we run when all the services are executed in parallel in our cluster. In particular, Figure~\ref{fig:roma-comparison}a shows the performance obtained by our framework, while Figure~\ref{fig:roma-comparison}b depicts the results obtained by a competitor. \roma was able to keep the response time of all the ML-services under the set SLAs (maximum response time equal to 0.4 seconds) thanks to its efficient usage of GPUs and vertical scaling of CPUs. The competitor solution frequently violated the SLAs of services \textit{VGG16} and \textit{ResNet} with a maximum response time of  $1.9$ seconds and an average resource allocation that is slightly higher compared to the one consumed by \roma.

%% file: conclusions.tex
The management of web services that use ML to provide intelligent features poses new complex challenges. ML-services require a training phase (batch) to generate a model that can be used in inference mode to compute predictions on new inputs (interactive). Time and quality requirements are needed to obtain predictable performance and accurate predictions.

This paper describes a set of critical aspects that characterize the management of ML-services in both the training and inference phases. As a base for a future, integrated framework that can handle all these aspects simultaneously, we also describe some initial partial solutions that we developed recently. A preliminary evaluation of them shows the benefits of our approaches that can efficiently and automatically fulfill identified requirements.